\documentclass{aastex}
\usepackage{emulateapj5}

\def\rp{$r^\prime$}
\def\gp{$g^\prime$}
\def\ip{$i^\prime$}
\def\up{$u^\prime$}
\def\zp{$z^\prime$}

\slugcomment{To appear in \apj. Last revision 22-FEB-2001.}

\shorttitle{SDSS/RASS Cluster Lensing}
\shortauthors{Sheldon, et al.}

\begin{document}


\title{Weak Lensing Measurements of 42 SDSS/RASS Galaxy Clusters}

\author{
Erin Scott Sheldon\altaffilmark{1},
James Annis\altaffilmark{2},
Hans B\"{o}hringer\altaffilmark{3},
Philippe Fischer\altaffilmark{4},
Joshua A. Frieman\altaffilmark{2,5},
Michael Joffre\altaffilmark{5},
David Johnston\altaffilmark{5},
Timothy A. McKay\altaffilmark{1},
Christopher Miller\altaffilmark{6},
R. C. Nichol\altaffilmark{6},
Albert Stebbins\altaffilmark{2},
Wolfgang Voges\altaffilmark{3},
Scott F. Anderson\altaffilmark{7},
Neta A. Bahcall\altaffilmark{8},
J. Brinkmann\altaffilmark{9},
Robert Brunner\altaffilmark{10},
Istv\'an Csabai\altaffilmark{11,12},
Masataka Fukugita\altaffilmark{13},
G. S. Hennessy\altaffilmark{14},
\v{Z}eljko Ivezi\'{c}\altaffilmark{8},
Robert H. Lupton\altaffilmark{8},
Jeffrey A. Munn\altaffilmark{15},
Jeffrey R. Pier\altaffilmark{15},
Donald G. York\altaffilmark{5,16}
}

\altaffiltext{1}{University of Michigan, Department of Physics,	500 East University, Ann Arbor, MI 48109}
\altaffiltext{2}{Fermi National Accelerator Laboratory, P.O. Box 500, Batavia, IL 60510}
\altaffiltext{3}{Max-Planck-Institut f\"{u}r Extraterrestrische Physik, D-85740 Garching, Germany}
\altaffiltext{4}{Deptartment of Astronomy, Univ. of Toronto, Toronto, ONT, M5S, 3H8, Canada}
\altaffiltext{5}{The University of Chicago, Department of Astronomy and Astrophysics, 5640 S. Ellis Ave., Chicago, IL 60637}
\altaffiltext{6}{Dept. of Physics, Carnegie Mellon University, 5000 Forbes Ave., Pittsburgh, PA-15232}
\altaffiltext{7}{University of Washington, Department of Astronomy, Box 351580, Seattle, WA 98195}
\altaffiltext{8}{Princeton University Observatory, Princeton, NJ 08544}
\altaffiltext{9}{Apache Point Observatory, P.O. Box 59, Sunspot, NM 88349-0059}
\altaffiltext{10}{Department of Astronomy, California Institute of Technology, Pasadena, CA 91125}
\altaffiltext{11}{Department of Physics and Astronomy, The Johns Hopkins University, 3701 San Martin Drive, Baltimore, MD 21218}
\altaffiltext{12}{Department of Physics of Complex Systems, E\"otv\"os University, P\'azm\'any P\'eter s\'et\'any 1}
\altaffiltext{13}{University of Tokyo, Institute for Cosmic Ray Reserach, Kashiwa, 2778582, Japan}
\altaffiltext{14}{U.S. Naval Observatory, 3450 Massachusetts Ave., NW, Washington, DC  20392-5420}
\altaffiltext{15}{U.S. Naval Observatory, Flagstaff Station, P.O. Box 1149, Flagstaff, AZ  86002-1149}
\altaffiltext{16}{The University of Chicago, Enrico Fermi Institute, 5640 S. Ellis Ave., Chicago, IL 60637}

\begin{abstract}
We present a lensing study of 42 galaxy clusters imaged in Sloan Digital Sky
Survey (SDSS) commissioning data. 
Cluster candidates are selected optically from SDSS imaging data and confirmed
for this study by matching to X--ray sources found independently in the ROSAT
all sky survey (RASS). 
Five color SDSS photometry is used to make
accurate ($\Delta z$=0.018) photometric redshift estimates that
are used to rescale and combine the lensing measurements. The 
mean shear from these clusters is detected to 2$h^{-1}$Mpc at the 7-$\sigma$
level, corresponding to a mass within that radius of
$(4.2 \pm 0.6) \times 10^{14}h^{-1}$M$_{\sun}$. The shear profile
is well fit by a power law with index $-0.9 \pm 0.3$, consistent with
that of an isothermal density profile. 
Clusters are divided by X--ray luminosity into two subsets, with mean 
L$_X$ of $(0.14\pm0.03)\times10^{44}$ and $(1.0\pm0.09)\times10^{44}h^{-2}$ergs/s.
The average lensing signal is converted to a projected mass density based
on fits to isothermal density profiles. From this we calculate a mean r$_{500}$
(the radius at which the mean density falls to 500 times the
critical density) and M$(<r_{500})$. 
The mass contained within
$r_{500}$ differs substantially between the low- and high-L$_X$ bins, 
with $(0.7\pm0.2)\times10^{14}$ and 
2.7$^{+0.9}_{-1.1}\times 10^{14}h^{-1}$M$_{\sun}$ respectively. 
This paper demonstrates our ability to measure ensemble cluster 
masses from SDSS imaging data.    
The full SDSS data set will include $\gtrsim$1000
SDSS/RASS clusters. With this large data set we will measure the M-L$_X$
relation with high precision and put direct constraints on the mass density of
the universe.
\end{abstract}

\keywords{dark matter --- galaxies: clusters: general --- gravitational lensing --- 
 	  large-scale structure of the universe --- X--rays: general}

\section{Introduction}\label{intro}
Galaxy clusters are massive, easily identifiable systems which can
be studied with a rich variety of techniques.
Scaling relations between their various measured properties reveal
important details of clusters physics.
The relation between the total mass and visible or X--ray luminosity
provides insight into the relative amount of ordinary and dark matter
in the cluster. Making the fair sample hypothesis, these scaling relations 
can be extrapolated to provide constraints on the cosmic density 
parameter, $\Omega_M$.

The total mass of a cluster is typically determined from the velocity dispersion of
its galaxies, or the X--ray emission profile and temperature of the hot
intracluster gas.
An alternative estimate of the total mass is provided by weak gravitational lensing.
Weak lensing has been used successfully to measure the mass of individual
clusters of galaxies \citep{Squires96,Smail97,Fischer97,Joff00}.  
Weak lensing measurements require deep images, providing a high density of
background objects.
The imaging data from the Sloan Digital Sky Survey 
(SDSS, see \cite{Fukugita96,Gunn98,York00}) is too shallow to 
measure the lensing signal
from individual clusters with high precision.  The large area of the 
survey, however, makes it ideal for measuring ensemble 
properties such as mean mass. These mean masses can be 
compared with any richness parameter; e.g. galaxy number density, 
optical luminosity, X--ray luminosity, etc, to reveal scaling relations.
An analogous measurement of the ensemble properties of galaxy halos using
SDSS data was reported in \cite{Fischer00} (F00).

There is a large effort to identify clusters in the
SDSS imaging data \citep{Annis00, Kim00, Nichol00}.
These studies will provide large, 
homogeneously selected
cluster samples containing thousands of new clusters. They will 
provide an excellent opportunity to measure cluster scaling
relations. As an initial exercise, clusters selected from SDSS commissioning
data using the method of \cite{Annis00} have been  
matched to sources in the Rosat All Sky Survey (RASS, see \cite{Voges99}).
This preliminary data set provides an opportunity to compare cluster
masses and X--ray luminosities in a statistical way.
Although this sample of 42 clusters is too small
to meaningfully constrain the L$_X$-M relation, we detect a clear correlation 
between lensing mass and X--ray luminosity. With the full SDSS data set we will 
measure the L$_X$-M relation with high precision.

In \S \ref{obs} we
outline the observations and cluster identification, and in \S \ref{lensanal}
we discuss the techniques used in the lensing analysis.  In \S \ref{results} we
present the basic measurements and fits to cluster models. Possible
sources of systematic error are addressed in \S \ref{systematics}. 
Conclusions and future work are discussed in \S
\ref{conclude}. Throughout this paper we use H$_0$ = 100 h km/s, and
assume a Friedman-Robertson-Walker cosmology with $\Omega_M$ =
0.3, $\Omega_{\Lambda}$ = 0.7.

\section{Observations}\label{obs}

\subsection{Imaging Data and Reduction} \label{lensobs}
Cluster selection and lensing analysis were performed on 225 square degrees
of Sloan Digital Sky Survey commissioning data  
obtained March 20-21 1999 (runs 752 and 756), 
and another 170 square degrees of commissioning data 
taken September 19 and 25 1998 (runs 94 and 125).  
The data include drift scan imaging in the
5 SDSS filters, \up, \gp, \rp, \ip, \zp, to a limiting magnitude of \rp = 23.0,
with seeing ranging from 1$^{\prime\prime}$ to 2$^{\prime\prime}$. 
The lensing study was performed on the \gp, \rp, and \ip \ images only, as the
\up\ and \zp\ bands are comparatively less sensitive.  The galaxy-galaxy
lensing study of F00 also uses the March 1999 data set, and more
details about the observations can be found there.

The SDSS photometric pipeline supplies fully calibrated photometric and
astrometric data for each object, along with a wealth of other parameters
including shapes \citep{Lup00}.  Because high S/N shape determination is a priority
for weak lensing studies (see \S \ref{lensanal}), we independently measure
the shape of each object. We
find the quadratic moments of the light distribution, weighted by an elliptical
gaussian adapted to the size and shape of the object through an iterative
process \citep{Bern00}:
\begin{equation}
Q_{i,j} = { {\sum I_{k,l} G_{k,l} x_i x_j } \over { \sum I_{k,l} G_{k,l} } },
\end{equation}
where $I_{k,l}$ is the sky subtracted surface brightness of the object at each
pixel \{k,l\} and 
$G_{k,l}$ is the elliptical gaussian weight.  
We then form the ellipticity
components from the quadratic moments:
\begin{equation}
e_1 = { {Q_{1,1} - Q_{2,2}} \over {Q_{1,1} + Q_{2,2}} }, \ \ e_2  =  
{ {2Q_{1,2}} \over {Q_{1,1} + Q_{2,2}} }.
\end{equation}
 The PSF is then characterized from the $e_1$ and
$e_2$ of stars, and the galaxy shapes are corrected for PSF anisotropy and
blurring as outlined in F00.

\subsection{Cluster Identification} \label{cluster}
\begin{figure}
\plotone{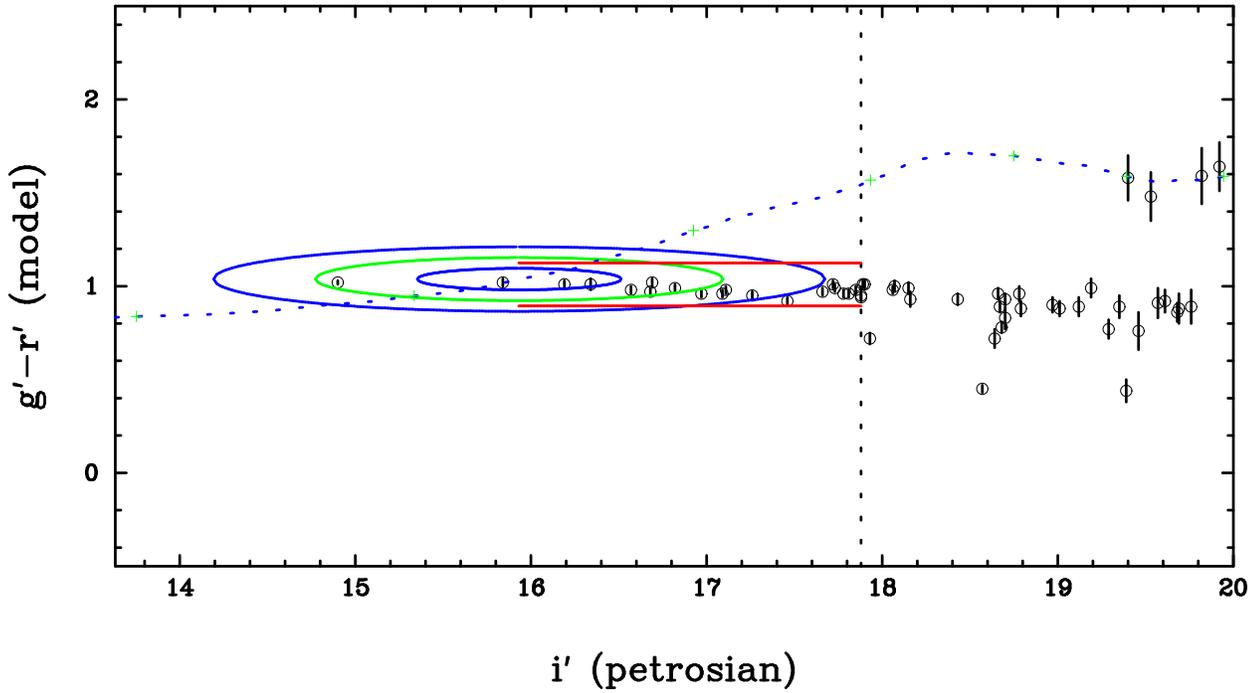}
\figcaption{ 
Color magnitude diagram for galaxies near a SDSS cluster at
redshift z=0.15.  
The dotted curve running from lower left to upper right is
the mean BCG color-magnitude relation as a function of increasing redshift.
Ellipses represent the
1-, 2- and 3-$\sigma$ contours around the mean BCG color and magnitude relation 
(determined from known BCG's measured in SDSS data), evaluated at the
best-fit redshift.
The vertical dotted line represents the 14
cutoff at 0.5$L_{*}$ in \ip\ at the redshift of the cluster.  
Galaxies in the ridgeline (defined by the 0.5$L_{*}$ line and the 2
horizontal lines running from \ip = 16 to 18) and 
within 0.667 Mpc of the BCG are counted as in the cluster. The number of
cluster members combined with a fit to the BCG color-magnitude relation
are used to estimate the likelihood of the cluster identification.
\label{bcg_color_mag} }
\end{figure}

The MaxBCG cluster identification algorithm \citep{Annis00}
relies on the fact
that brightest cluster galaxies (BCGs) form a narrowly defined class; they
have a small dispersion in luminosity and an even tighter relation in color
(see also \cite{Gladd00}).  For each SDSS galaxy we calculate a 
``BCG likelihood'' based on its color and magnitude. 
We combine this BCG likelihood with the number of
neighboring galaxies found in the corresponding E/SO ridge line to define a cluster
likelihood. This likelihood is 
calculated for every redshift from 0.0 to
0.6 at intervals of 0.01. The redshift is chosen to maximize the cluster likelihood.
Elliptical galaxies, which have very regular colors, provide excellent
photometric redshift targets. Since each cluster includes a number of these
galaxies, photometric redshift estimates for these clusters are excellent.
The dispersion between 2406 estimated BCG redshifts and their SDSS spectroscopic
redshifts is $\Delta z$ = 0.018. Figure \ref{bcg_color_mag} shows a schematic
of the technique. 

\subsection{X--ray Matches} \label{xraydata}
\begin{figure}
\plotone{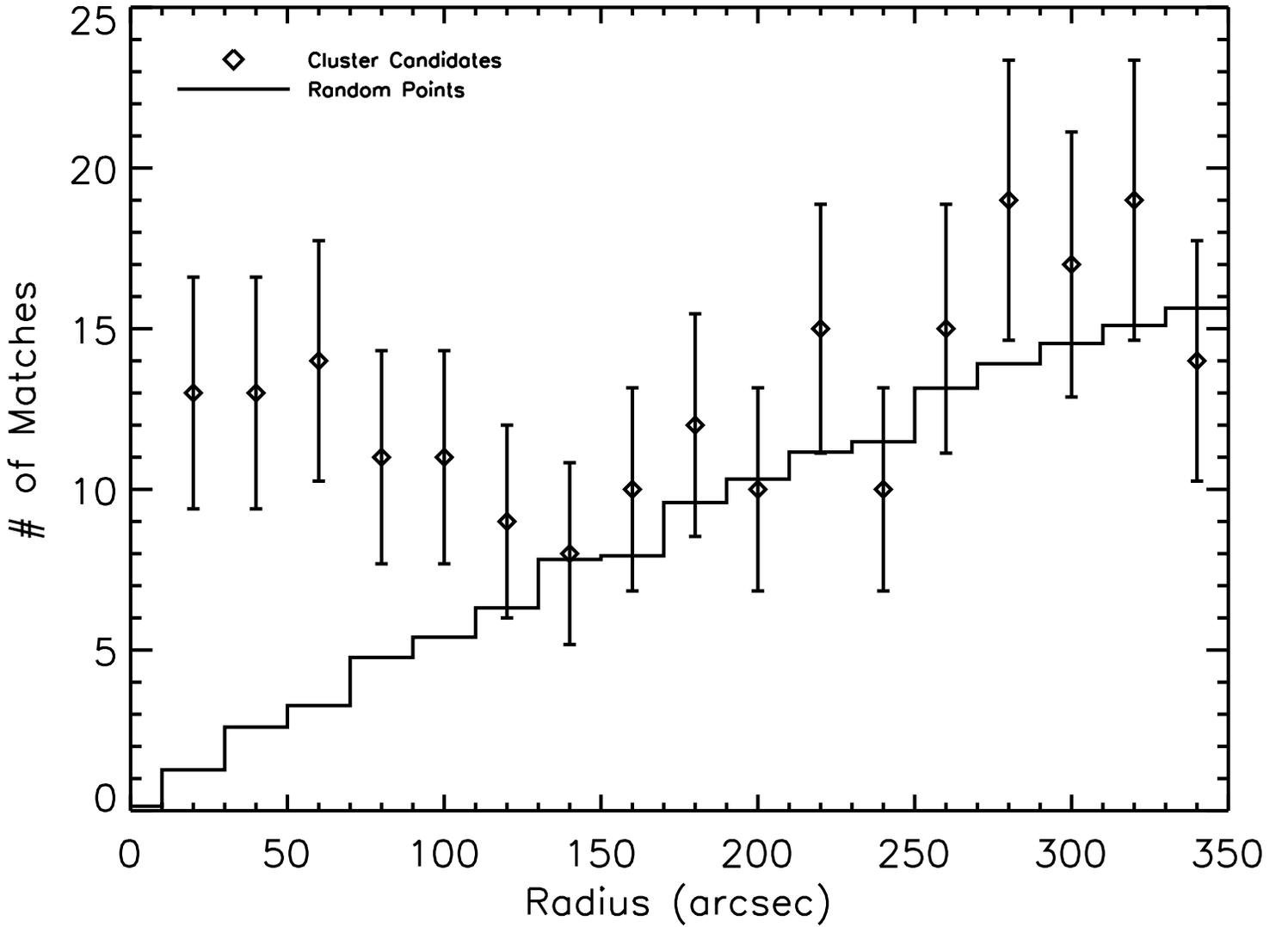}
\figcaption{Angular separations 
between optically selected cluster candidates
and RASS bright and faint source catalog objects.  The diamonds
are the number of matches between the MaxBCG objects and
RASS. Error bars are $\pm 1\sigma$ Poisson. The solid line is the mean number of matches to 100 sets of random points, each set
containing the same number of points as the MaxBCG. The ratio of the cluster
matches to the random matches is the purity of the sample at a given radius. 
The total purity within 80$^{\prime\prime}$ is $\sim$ 75\%. \label{matches} }
\end{figure}

The X--ray data are taken from the ROSAT All-Sky Survey 
\citep{Truemp83, Voges99}.  
The combined RASS bright and faint source catalogs \citep{Voges99,Voges00_2}
are correlated with the MaxBCG sources discussed above to find
X--ray emitting 
SDSS/RASS clusters (SRC). Matches were
found based on a search radius of 80$^{\prime\prime}$ which corresponds to a
net purity of 75\% based on comparison with random associations (see Figure
\ref{matches}). Initial optical selection allows us
to identify X--ray emitting clusters at fluxes substantially below the flux
limits of X--ray selected cluster surveys. This sample was then
visually inspected and potentially contaminated cluster matches were excluded
for the analysis discussed in this paper. This process yielded a conservative sample
of 50 SRC clusters.

As the X--ray count rate for extended sources like clusters of galaxies in the
RASS is not very accurate, we used both the Growth Curve Analysis of 
\cite{Bohr00} and an independent analysis to compute the count rates of these
clusters. Both techniques agree within the uncertainties.  
These count rates are converted to 
bolometric X--ray luminosities
using the photometric redshift, a luminosity--temperature relationship, 
and the ROSAT response matrix. 
Details are reported in \cite{Voges01}.

\section{Lensing Analysis}\label{lensanal}
Foreground mass distributions induce coherent distortions in the images
of background galaxies.  On average, background galaxies appear 
tangentially aligned with respect to the foreground lens.  
The tangential shear, $\gamma_+$,
is related to the projected mass
distribution of the foreground lens, $\Sigma$, by
\begin{equation} \label{sheareq1}
\gamma_+(R) = \overline{\kappa}(\le R) - \overline{\kappa}(R),
\end{equation}
%
where $\kappa = \Sigma/\Sigma_{crit}$ is the surface density normalized by 
$\Sigma_{crit}$, the critical
density for multiple lensing \citep{Escude91}.
The first term on the 
right hand side of equation \ref{sheareq1} is the mean scaled density interior to projected
radius $R$ and the second is the mean scaled density at $R$. The critical density 
depends only on the geometry of the source-lens-observer system:
\begin{equation} \label{sigcriteq}
\Sigma_{crit} = { c^2 \over {4\pi G} } {D_S \over {D_L D_{LS}} },
\end{equation} 
where $D_S$ and $D_L$ are the angular diameter distances from observer to to 
source and lens,
respectively, and $D_{LS}$ is the angular diameter distance from lens to source.

Assuming that the orientations of unlensed background galaxies are uncorrelated with the 
position of a foreground mass, the shear can be
estimated directly from the shapes of background galaxies.  It can be
shown that the shear induced by weak lensing ($\kappa << 1$) is simply equal to
half the induced ellipticity as measured in the tangential frame \citep{Escude91}.  
We denote the tangential component of the ellipticity as $e_+$, and the
orthogonal component as $e_{\times}$.  Note the ellipticity parameter
transforms as 2$\theta$ under rotations, so the ``orthogonal'' frame is
rotated by 45$\degr$ with respect to the tangential. 

Galaxies are intrinsically elliptical, so
the shear measurement must be done statistically by averaging the shapes of many
source galaxies:
\begin{equation}
\hat{\gamma}_+ = {1 \over {2 S_{Sh}} } \langle e^i_+ \rangle = 
{1 \over {2S_{Sh}}}{ {\sum{w_i e^i_+} \over {\sum{w_i}}} },
\end{equation}
where $S_{Sh} \sim 0.88$ is the average responsivity of the source galaxies to a shear 
as defined in F00, and is similar to the shear polarizability defined in 
\cite{ka95}. The $w_i$ are the weights for each shape measurement.

%

There are two dominant sources of random noise in shear measurements: the intrinsic
variance in galaxy shape, or ``shape noise'' 
$\sigma_{SN} = \langle e_i^2 \rangle \sim 0.32$, and the uncertainty
in the shape measurement of each galaxy $\sigma_i \sim 0.25$. 
For optimal S/N we weight by $1/N^2$, $w_i = 1/(\sigma_i^2. + \sigma_{SN}^2)$.

Because SDSS images are relatively shallow, the number density of source
galaxies behind each lens is low, about 2 per square arcminute.
For this cluster sample, the typical mean induced ellipticity within 600$^{\prime\prime}$
is $2\gamma_+ < 0.01$.
Thus our sensitivity, $\sim 0.4/\sqrt{N_{Source}}$, is
insufficient to detect the shear induced by a single lens with precision. 
To improve the S/N, we average the shear from many lenses.  

Averaging over many lenses increases the signal to noise of the shear
measurement. Each lens has a different lensing strength, however, which scales as
$\Sigma_{crit}^{-1}$. This means lenses with the same mass but different
redshifts will not produce the same shear.  We rescale the shear to a density
contrast, which is a redshift independent quantity, by multiplying both sides
of equation \ref{sheareq1} by $\Sigma_{crit}$:
\begin{equation}
\Delta\Sigma \ \tbond \ \overline{\Sigma}(\le R) - \Sigma(R) = \gamma_+ \Sigma_{crit}.
\end{equation}
The weight for each measurement, $w_i$, must also be rescaled
to $W_i = w_i\Sigma_{crit}^{-2}$.

\begin{figure}
\plotone{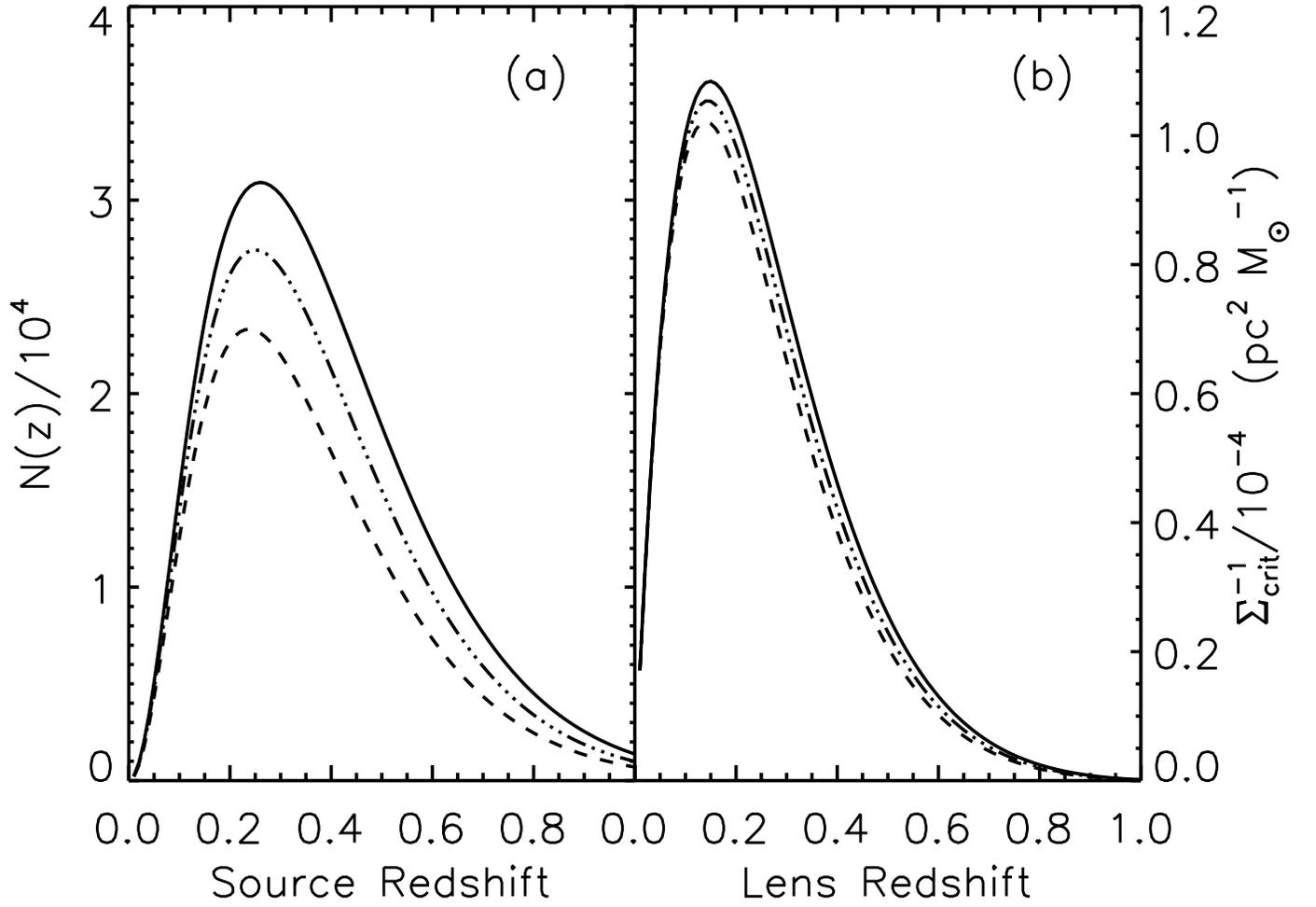}
\figcaption{(a) Redshift distributions for source
galaxies found in \gp (dashed line), \rp (solid), and \ip (dot-dashed) inferred from their
\rp magnitudes. (b) The lensing strength, $\Sigma_{crit}^{-1}$, vs. lens redshift,
calculated from the redshift
distributions shown in (a).\label{nzscrit}}
\end{figure}

The total weight for a given cluster is $\sim N_{source}/\Sigma_{crit}^2$, 
assuming that the measurement
error is approximately the same for each source. Because the lensing signal is 
proportional to $1/\Sigma_{crit}$, this is equivalent to weighting by
the relative $\left(S/N\right)^2$. Nearby clusters will have larger
$N_{source}$, whereas clusters at $z \sim 0.15$ will have 
large lensing strength $1/\Sigma_{crit}$ (see figure \ref{nzscrit}(b)).  The
combination of these two effects conspires to give
clusters at $z\sim0.09$ the largest relative weight.

To find the average density contrast, we need to determine
$\Sigma_{crit}^{-1}$, the lensing strength of each lens-source pair, which
depends on the redshift of both lens and source. 
We have a good photometric redshift for each cluster, but we do not have a 
redshift for each source
galaxy. As a result, we estimate the average lensing strength of each cluster by
%
integrating over the source redshift distribution:
\begin{equation} \label{meanscrit}
\Sigma_{crit}^{-1}(z_L) = { {\int_{z_L}^{\infty} \Sigma_{crit}^{-1}(z_L,z_s) P(z_s) dz_s} },
\end{equation}
where P($z_s$) is the normalized source redshift distribution. The fact that
some source galaxies are in front of the lens, and dilute the signal, is
accounted for by beginning the integral at $z_L$.

To estimate the redshift distribution in equation \ref{meanscrit} 
we first gather a sample of galaxies with
known redshifts 
drawn from the SDSS and the Canada-France Redshift 
Survey
(CFRS: \cite{Lil95}).  CFRS galaxies have magnitudes measured in V and I.  We
convert these to \rp\ using the relation $r^{\prime} = I_{CFRS} + 0.5(V_{CFRS} - I_{CFRS})$.
This relation is empirically determined from galaxies observed by both SDSS
and CFRS.

The redshift distributions of these galaxies, in bins of \rp, are fit to the
function (\cite{Baugh93})
\begin{equation} \label{zfit}
n(z) \varpropto z^2 exp(-({ {z}/ {z_c} })^{3/2}).
\end{equation}
The values of $z_c$ are then used to derive a relationship between \rp\ and $z_c$.
The derived $z_c$ vs. \rp \ relation has been independently confirmed
by comparison of SDSS and CNOCII data \citep{Yee00}.

For the lensing measurements, we choose ``background,'' or source galaxies with 
reddening corrected Petrosian magnitude
18.0 $<$ \rp $<$ 22.0 and size 4$\sigma$ larger than the stellar locus (see
F00). We have source galaxy catalogs with measured
shapes in \gp, \rp, and \ip \ bands, containing 
1.7, 2.4, and 2.1 million
objects respectively. The magnitude distribution of each catalog is converted
to a redshift distribution using the derived \rp-$z_c$ relationship.
The inferred redshift distribution for each of these catalogs is shown in
Fig. \ref{nzscrit}(a), and \ref{nzscrit}(b) shows 
the lensing strength $\Sigma_{crit}^{-1}(z_{L})$
derived from the redshift distributions.

Some source galaxies are in front of the lens, and this is accounted for
in equation \ref{meanscrit}, but
a few galaxies in our source catalog are physically
associated with the clusters. The number density of these galaxies will
decrease with physical radius from the center of the cluster and  
their inclusion in the analysis will radially bias the shear estimate. 
To correct for this effect, we compare the number of background galaxies around 
random points to those around 

lenses. We choose 100 random points for each lens,
and assume an excess of background galaxies is due to objects 
physically associated with the
lens. We correct for this
dilution by multiplying the measured shear by a radial correction factor 
F(r) = $\langle N_{lens}\rangle/\langle N_{rand}\rangle$, the ratio
of source galaxies around lenses to that around random points.
The average correction factor ranges from a factor of 1.5 in the 
innermost bin to 1.02 in the outermost (see \S \ref{profile}). 

It should be noted that the magnification of background sources by 
the foreground lens may also increase the 
number density of neighboring 
galaxies. The magnification will bring faint galaxies into our magnitude
limited catalog. On the other hand, the geometric distortion introduced by the
lens moves background galaxies apart, diluting the number density of sources.  
The relative change in the number density goes as 
$N^{\prime}/N_{0} \varpropto \mu^{2.5s-1}$, where $\mu = 1+\epsilon$ is the 
magnification and s is the slope of the galaxy number 
counts \citep{Broad95}. For these clusters, 
$\epsilon \sim$ shear $\lesssim .01$. Thus, for s on the order of 0.4,  
which it is for our background galaxies, 
the effect of magnification is small compared to the neighbor excess 
which is $\lesssim 0.5$.

\section{Results}\label{results}
\begin{figure}
\plotone{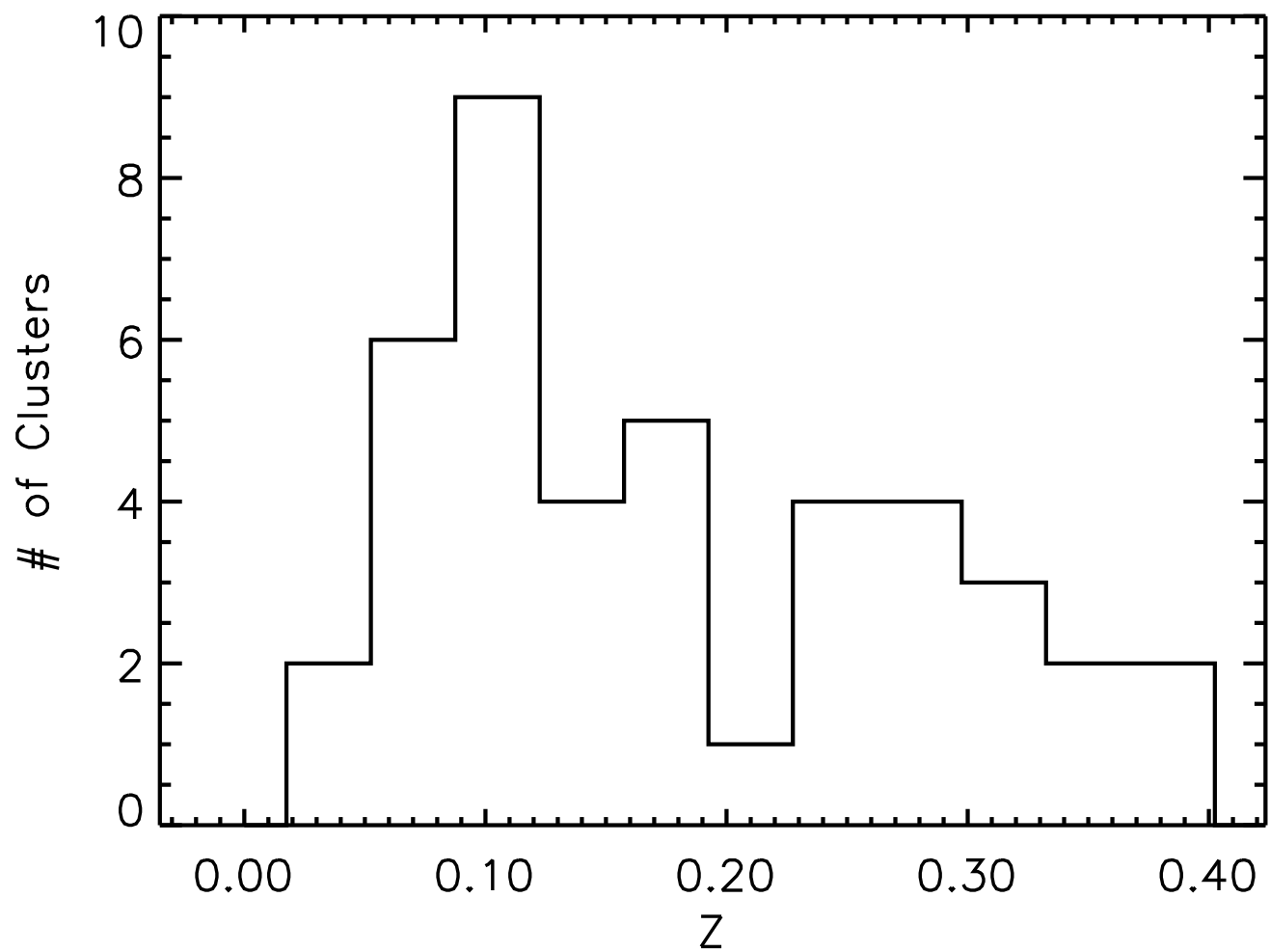} 
\figcaption{Photometric 
redshift distribution for the 42 SRC clusters used in the lensing analysis.
\label{lenszhist}}
\end{figure}

\begin{figure}
\plotone{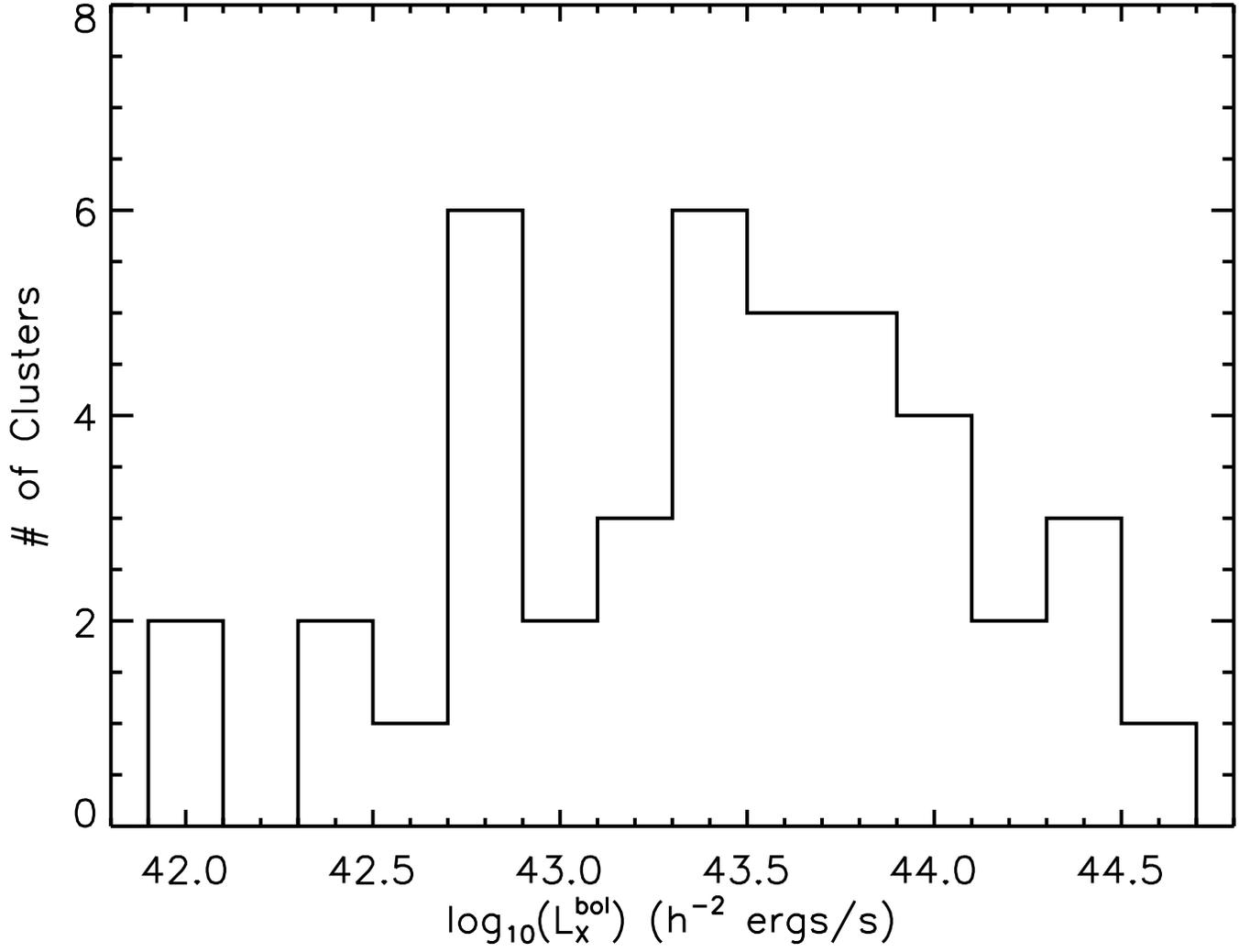} 
\figcaption{Distribution of
bolometric X--ray luminosities for the 42 SRC clusters used in the lensing
analysis.  \label{lxhist}}
\end{figure}

We begin with 50 clusters found in the SDSS imaging data and matched to 
RASS as described in \S \ref{xraydata}.
We then choose clusters with redshift $z < 0.4$. The
upper limit of z=0.4 is where photometric redshift estimation becomes less
accurate as the 4000\AA\  break passes between the \gp\ and \rp\ bandpasses. 
To ensure that residual systematic shape
correlations do not bias the shear
estimate, we demand that the azimuthal distribution of source galaxies around each
lens be reasonably symmetric. After these cuts we are left with 42
clusters, with a photometric redshift distribution is shown in 
Figure \ref{lenszhist}. Their ROSAT bolometric X--ray luminosities range from
8.5$\times10^{41}$ to 4.4$\times10^{44}$ ergs/s, with a distribution of X--ray
luminosities as shown in Figure \ref{lxhist}

\subsection{Cluster Density Profile}\label{profile}
\begin{figure}
\plotone{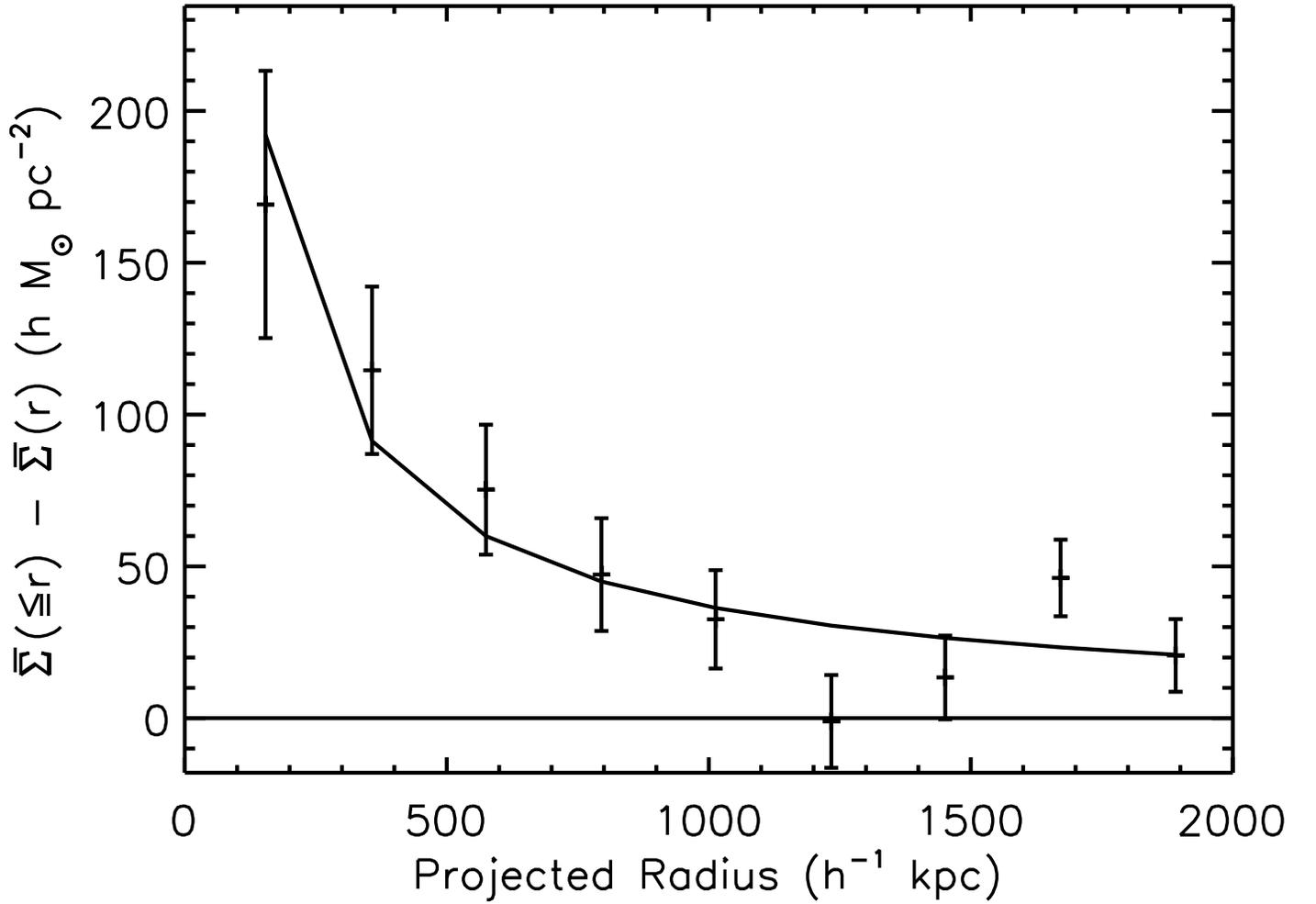}
\figcaption{Average projected density contrast for 42
SRC clusters. For this plot we have combined the measurements made 
in the \gp, \rp, and \ip \ bandpasses.  The
solid line is the best fitting power law, $\Delta\Sigma \varpropto
R^{-0.9}$. Errors bars are $\pm 1\sigma$. \label{subz_dens_cont}}
\end{figure}

The shear in the annulus 20-2000 kpc, centered on the BCG, 
is $(3.3 \pm 0.5)\times10^{-3}$, which
is a detection at the $\sim$7-$\sigma$ level (see Table \ref{table1}). 
Because the shear is a redshift dependent 
quantity, we rescale each shear measurement 
%
%
%
to a density contrast
as discussed in \S \ref{lensanal}.
Fig. \ref{subz_dens_cont} shows the average density contrast measured in nine
radial bins centered on the BCG.
The independent radial bins have centers ranging from 150 to 1890 $h^{-1}$ kpc.
We have applied a correction 
factor for contamination by cluster members as described in \S \ref{lensanal} 
For this figure we have combined the measurements made 
in the \gp, \rp, and \ip \ bandpasses.  

As a test to see if our signal is due to gravitational lensing, 
we have measured the orthogonal component of the shear, 
$\gamma_{\times} = \langle e_{\times}\rangle/2$ in the same radial
bins used for measurement of the tangential shear.  
This measurement is equivalent to measuring the curl of the gradient of
$\kappa$ and 
should be zero \citep{ka95a,Stebbins96}.  
We find that the $\gamma_{\times}$ profile is indeed a good fit to zero 
($\chi^2$ = 7.0/9).
We have also run tests for systematics
in the background galaxy shapes and bias in the shear measurement due to the faint tails
of foreground galaxy light distributions. Both of these tests are null and
details can be found in F00.

We fit the observed density contrast in \gp, \rp, and \ip \ to a power 
law of the form 
$\Delta\Sigma(R) \varpropto R^{\gamma}$,
using the full covariance matrix to estimate the chi squared statistic.  We find 
$\gamma = -0.9 \pm 0.3$.  A power law shear profile implies 
the same power law for the projected mass density profile.   Note again that we
have used the BCG as the center for our measurements. The use
of any position for the center which is not the true center of mass will
complicate the interpretation of these fits (see \S \ref{systematics}).

\subsection{Cluster Model}\label{mass}
The average cluster density contrast is consistent with that of a projected 
singular isothermal
sphere (SIS); $\gamma = -1$. 
Cluster density profiles inferred from lensing
are typically well fit by an isothermal \citep{Fischer97}.  We will use
this model for the remainder of the paper to represent the average cluster
density profile.
The SIS model has projected density
\begin{equation}\label{sigmasis}
\Sigma(R) = 116 \left( \sigma_v \over 1000 \textrm{km/s} \right)^2 
\left(R \over 1\textrm{Mpc}\right)^{-1} \textrm{M$_{\sun}$ pc$^{-2}$}.
\end{equation}
To estimate masses, we simply fit the measured profile to a 1/R density with one 
free parameter, the velocity dispersion $\sigma_v$.  Note that a SIS has a 
density contrast equal to the density itself, $\Delta\Sigma(R) = \Sigma(R)$.  



Once the velocity
dispersion is estimated, the density can be integrated to give the total mass
within a projected radius $R$.
The mass of an SIS within projected radius R is a factor of $\pi/2$
larger than the mass within the three dimensional radius r=R.

\subsection{Mass vs. L$_X$} \label{masslx}

Table \ref{table1} contains fits to equation \ref{sigmasis} for 
the sample as a whole as well as
two subsamples, split by their X--ray luminosity.  The shear profiles for the 
subsamples are 
both consistent with isothermal. The average bolometric L$_x$
is calculated by weighting the
individual L$_x$ of each 
cluster with its relative weight in the shear measurement $\approx N_{source}/ \Sigma_{crit}^{2}$.

We use the fits to equation \ref{sigmasis} to infer the velocity dispersion 
and the mass within 1 Mpc. The mass of the higher L$_X$ subsample is
substantially higher than the low L$_X$ subsample. Because the clusters are 
different sizes, it is more meaningful to discuss the mass within a 
characteristic radius $r_{500}$, the radius within which the mean density falls
to 500 times the critical density at that redshift. The fits 
to $r_{500}$ and the 
projected mass within $r_{500}$ are also shown in Table \ref{table1}. Again, 
the mass of the higher L$_X$ subsample is substantially higher.



\section{Systematic Error} \label{systematics}
Possible sources of systematic error in the mass measurements are 
errors in the
correction for blurring by the 
PSF and inclusion of stars in the source galaxies.
The blurring correction and stellar contamination were addressed in F00, and are
less than 10\% and 1\% respectively.
Other possibilities are variations of angular diameter distance with
cosmological model, unertainties in the source galaxy redshift 
distribution,
and offsets of the BCG from the true center of mass of the
cluster. 

Lensing masses were calculated in 3 different cosmologies: 
$\left(\Omega_M, \Omega_{\Lambda}\right)$ = $\left(1.0,0.0\right)$, 
$\left(0.3,0.0\right)$, $\left(0.3,0.7\right)$.
We find that the mass estimates vary by less than 5\% between these
cosmologies.

The procedure for
estimating the redshift distribution is based on the \rp\  
magnitude distribution of the sources 
%
(see \S \ref{lensanal}).
In order to understand the uncertainty in the derived distribution, 
we examined the range of allowed 
$z_c$ values in the fit to Eq. \ref{zfit}.
Using the 68\% confidence limits from our
$\chi^2$ fits, we find that the derived masses vary by less than 5\%.

Differences in selection criteria between galaxies in CFRS and SDSS might also
bias the assumed source redshift distribution.
We intend to address this limitation in the future by directly obtaining
redshifts for a sub-sample of the source galaxy catalog.

In this paper we have used the BCG as the center for our lensing measurements.
As we have stated above, the use of any center other than the true center of
mass may 
bias the mass estimates. 
An alternative is to use the RASS position, but these are ill-determined for
these low-flux, extended sources \citep{Voges00_2}. The mean offset between BCG and 
measured X--ray centroid is 85kpc (max $\sim$200kpc). However, because we use 
apertures much greater than 85kpc, we do not expect our mass measurements to differ 
significantly
for these centroid shifts. We have repeated our analysis using the
X--ray centroid as the center and the signal remained within the 
1-$\sigma$ error limits. 

\begin{figure}
\plotone{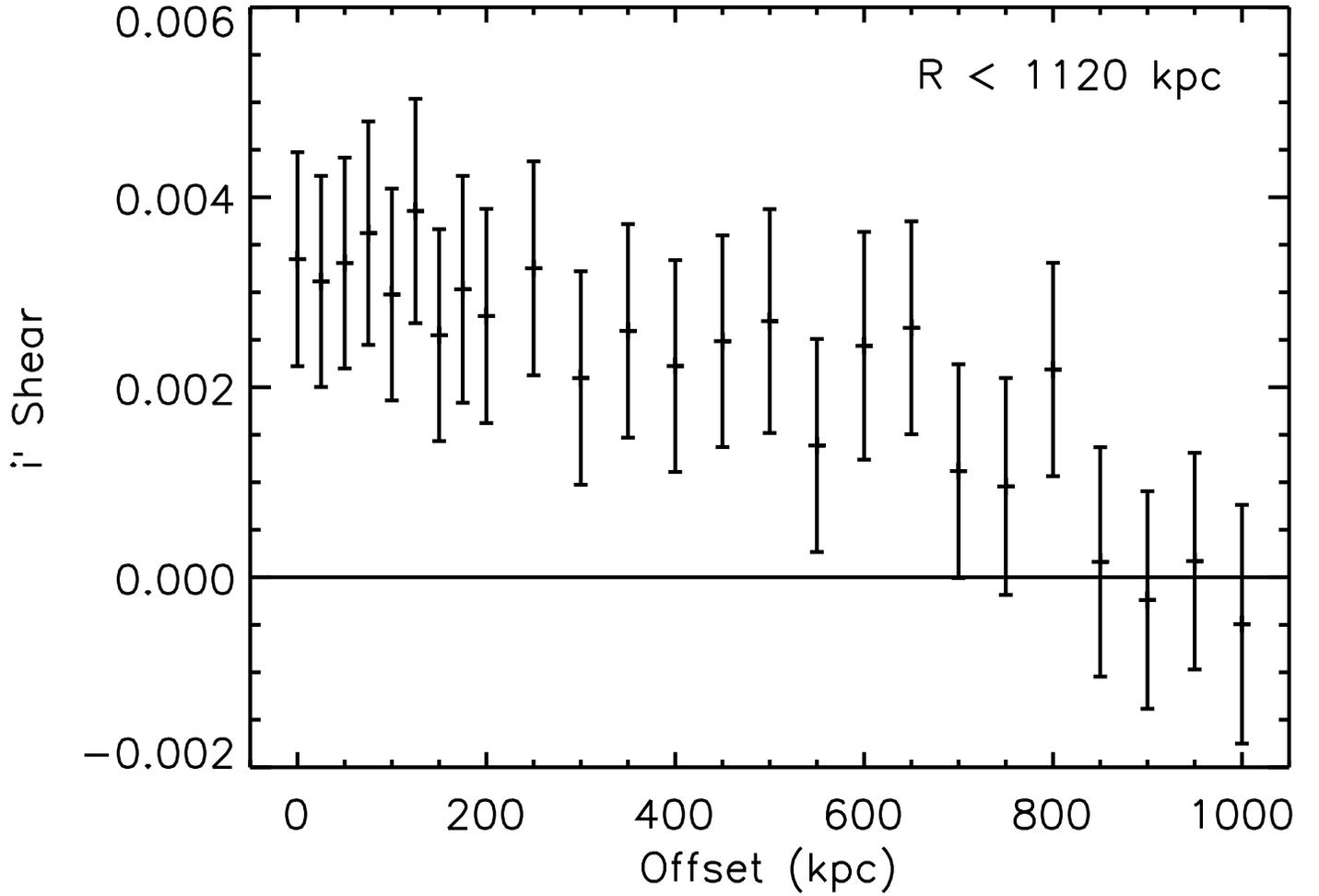} 
\figcaption{Shear within a 1.1Mpc 
aperture measured as a function of 
artificial offset from the BCG. Shear is measured in the \ip \ band. 
The shear is not changed substantially for
offsets less than 200kpc.  The shear becomes consistent with zero as the
offset approaches the size of the aperture. \label{rassoffset}}
\end{figure}

We have further checked our sensitivity to offsets
by repeating the analysis with centers artificially offset from the BCG by various
amounts and in random directions. Figure \ref{rassoffset} 
shows the \ip-band shear in a 1.1Mpc aperture measured as a function of artificial
centroid shift. The shear is not changed substantially for shifts less than
about 200kpc. 
Our mass estimates are also unchanged for shifts less than
200kpc. Furthermore, the shear becomes consistent with zero for 
shifts that approach the size of the aperture.  
Because we do not know the true mass distribution of these
objects, we cannot measure the offsets with this data alone. However, this
test is consistent with the assumption that the true center is within the
aperture and is less than 200kpc from the BCG.  Until we can find the true
center of mass, or at least estimate the mean size of the offsets, we
will not be able to quantitatively determine the effect of these offsets on 
our mass estimates. This issue will be dealt with in a future paper.

An issue related to centroid offsets is the problem of sub-structure   
within clusters. If there are substantial sub-clumps in a cluster, not
only is the BCG unlikely to be at the centroid, but the density profile 
will not be isothermal. In particular, the slope of the shear profile at
large radii will become flatter than expected from an isothermal, as
sub-clumps become included within the aperture \citep{Clowe00}. This will
make an SIS model an inappropriate representation of the data.
However, the reduced $\chi^2$ for the SIS fits to 42 clusters is  
3.3/8=0.4, suggesting that, in the ensemble average, sub-clumping is not
a dominant source of error.

\section{Conclusions}\label{conclude}

We have demonstrated the ability to measure ensemble cluster masses using SDSS
imaging data.  We detect the shear from 42 SRC
clusters with high signal to noise,
and find a correlation between the derived lensing mass and X--ray
luminosity.  The current sample represents only 4\% of the final SDSS data
set.  Extrapolating, we expect $\gtrsim$ 1000 MaxBCG/RASS clusters, and
perhaps more with the addition of the catalogs of \cite{Kim00} and \cite{Nichol00}.
With the final sample, we will
be able to measure the M-L$_X$ relation with high precision.  
The excellent 5-band photometry of the SDSS will yield measurements
of the total light in each cluster. Furthermore, SDSS spectroscopy of the nearby
clusters will yield an estimate their velocity dispersion. Thus in addition to
the L$_X$-M, we will
measure scaling relations between lensing mass, visible light in 5 bandpasses, 
and velocity dispersion, all measured from SDSS data. These scaling relations 
will greatly enhance our understanding of cluster physics.

Scaling relations for other classes of objects can be studied in a similar 
fashion.  Though individual galaxies and groups are much less massive than
clusters, and hence are less effective lenses, they are far more common. As a
result, our sensitivity to ensemble masses of galaxies (F00), groups, and
clusters are similar. This new ability to measure ensemble masses for classes
of objects should contribute substantially to our understanding of the
relationship between mass and luminous matter in the universe.

\acknowledgements
We thank Gary Bernstein for helpful discussions, especially
concerning shape measurements and corrections.

The Sloan Digital Sky Survey (SDSS) is a joint project of The University of 
Chicago, Fermilab, the Institute for Advanced Study,
the Japan Participation Group, The Johns Hopkins University, the 
Max-Planck-Institute for Astronomy, New Mexico State
University, Princeton University, the United States Naval Observatory, and 
the University of Washington. 
Apache Point
Observatory, site of the SDSS telescopes, is operated by the 
Astrophysical Research Consortium (ARC). Funding for the project has
been provided by the Alfred P. Sloan Foundation, the SDSS member institutions, 
the National Aeronautics and Space Administration, the National Science
Foundation, the U.S. Department of Energy, and Monbusho. The SDSS Web site
is http://www.sdss.org/. Tim Mckay acknowledges support through NSF PECASE
AST9703282. Bob Nichol acknowledges support through NASA LTSA grant NAG5-6548.

This research has made use of the NASA/IPAC
Extragalactic Database (NED) which is operated by the Jet Propulsion Laboratory,
California Institute of Technology, under contract with the National 
Aeronautics and Space Administration.

\clearpage

\clearpage

\begin{deluxetable}{ccccccc}
\tabletypesize{\small}
\tablecaption{Fits to Singular Isothermal Spheres \label{table1}}
\tablewidth{0pt}
\tablecomments{The first row is the average over the entire sample of clusters
and the subsequent rows are subsamples. All masses are projected values.}
\tablehead{
\colhead{$\langle L_X^{bol} \rangle$} & 
\colhead{N$_{Clust}$} & 
\colhead{$\langle \gamma_T(20-2000$kpc$) \rangle$} &
\colhead{$\sigma_v$} &
\colhead{M($< 1 h^{-1}$Mpc)} &
\colhead{$r_{500}$} &
\colhead{M($<$ $r_{500}$ )} \\
\colhead{($\times10^{44} h^{-2}$ ergs/s)} & 
&
\colhead{$( \times 10^{-3})$} &
\colhead{(km/s)} &
\colhead{($\times10^{14} h^{-1} $M$_{\sun}$)} &
\colhead{($h^{-1}$Mpc)} &
\colhead{($\times10^{14} h^{-1} $M$_{\sun}$)}
}
\startdata
0.30 $\pm$ 0.06 & 42 & 3.3$\pm$0.5 & 530$^{+40}_{-50}$  & 2.1$^{+0.3}_{-0.4}$ & 0.41 & 0.9 $\pm$ 0.2\\
0.14 $\pm$ 0.03 & 27 & 2.2$\pm$0.5 & 490 $\pm$ 60       & 1.8 $\pm$ 0.4       & 0.39 & 0.7 $\pm$ 0.2\\
1.0  $\pm$ 0.09 & 15 & 9.0$\pm$1.5 & 800$^{+90}_{-110}$ & 4.7$^{+1.0}_{-1.3}$ & 0.57 & 2.7$^{+0.9}_{-1.1}$\\
\enddata
\end{deluxetable}

\end{document}